\documentclass[lettersize,journal]{IEEEtran}
\usepackage{amsmath,amsfonts}
\usepackage{algorithmic}
\usepackage{array}
\usepackage{textcomp}
\usepackage{stfloats}
\usepackage{url}
\usepackage{verbatim}
\usepackage{graphicx}
\usepackage{cite}

\usepackage{subfigure}
\usepackage{makecell}
\usepackage{listings}
\usepackage{multirow}
\usepackage[ruled,linesnumbered]{algorithm2e}
\usepackage{array}
\usepackage{verbatim}
\usepackage{tablefootnote}

\newcommand{\mn}{PePNet}
\begin{document}

\title{A Heavy-Load-Enhanced and Changeable-Periodicity-Perceived Workload Prediction Network\thanks{$^\dagger$Corresponding author.}}

\author{\IEEEauthorblockN{Feiyi Chen$^1$, Naijin Liu$^2$, Zhen Qin$^1$, Hailiang Zhao$^1$, Mengchu Zhou$^3$, Shuiguang Deng$^1$$^\dagger$} \\
\textit{$^1$Zhejiang University} \quad
\textit{$^2$Beijing Wuzi University} \quad 
\textit{$^3$Zhejiang Gongshang University}\\
\textit{$^1$\{chenfeiyi,zhenqin,hliangzhao,dengsg\}@zju.edu.cn}
\textit{$^2$\ 2321350021@bwu.edu.cn}
\textit{$^3$\ mengchu@gmail.com}
}

% The paper headers
\markboth{Journal of \LaTeX\ Class Files,~Vol.~14, No.~8, August~2021}%
{Shell \MakeLowercase{\textit{et al.}}: A Sample Article Using IEEEtran.cls for IEEE Journals}

% Remember, if you use this you must call \IEEEpubidadjcol in the second
% column for its text to clear the IEEEpubid mark.

\maketitle

\begin{abstract}
Cloud providers can greatly benefit from accurate workload prediction. However, the workload of cloud servers is highly variable, with occasional workload bursts, which makes workload prediction challenging. The time series forecasting methods relying on periodicity information, often assume fixed and known periodicity length, which does not align with the periodicity-changeable nature of cloud service workloads. Although many state-of-the-art time-series forecasting methods do not rely on periodicity information and achieve high overall accuracy, they are vulnerable to data imbalance between heavy workloads and regular workloads. As a result, their prediction accuracy on rare heavy workloads is limited. Unfortunately, heavyload-prediction accuracy is more important than overall one, as errors in heavyload prediction are more likely to cause Service Level Agreement violations than errors in normal-load prediction.
  Thus, we propose a changeable-periodicity-perceived workload prediction network (PePNet) to fuse periodic information adaptively for periodicity-changeable time series and improve rare heavy workload prediction accuracy. It has two distinctive characteristics: 
  (i) A Periodicity-Perceived Mechanism to detect the periodicity length automatically and fuses periodic information adaptively, which is suitable for periodicity-changeable time series, and 
  (ii) An Achilles' Heel Loss Function that is used to iteratively optimize the most under-fitting part in predicting sequence for each step, thus evidently improving the prediction accuracy of heavy load.
  Extensive experiments conducted on real-world datasets demonstrate that PePNet improves accuracy for overall workload by 11.8\% averagely, compared with state-of-the-art methods. Especially, PePNet improves accuracy for heavy workload by 21.0\% averagely. 
\end{abstract}


\section{Introduction}
\IEEEPARstart{A}ccurateAccurate workload prediction brings huge economic benefits to cloud providers \cite{niknafs2019runtime} and many cloud frameworks make real-time adjustments based on the workload prediction results \cite{10.1145/3041021.3054186}.
On the one hand, workload prediction provides meaningful insights to improve the utilization of cloud servers while ensuring the Quality of Service (QoS) \cite{9209730} \cite{DBLP:conf/scc2/MoussaYBDH19,bi2019temporal,bi2024arima,yuan2024improved}.
On the other hand, it can alarm forthcoming rare heavy workload to avoid Service Level Agreement (SLA) violations.

\begin{figure}[t]
  \centering 
  \subfigure[The overall and heavy-workload prediction accuracy.]{
  \includegraphics[width=0.45\linewidth]{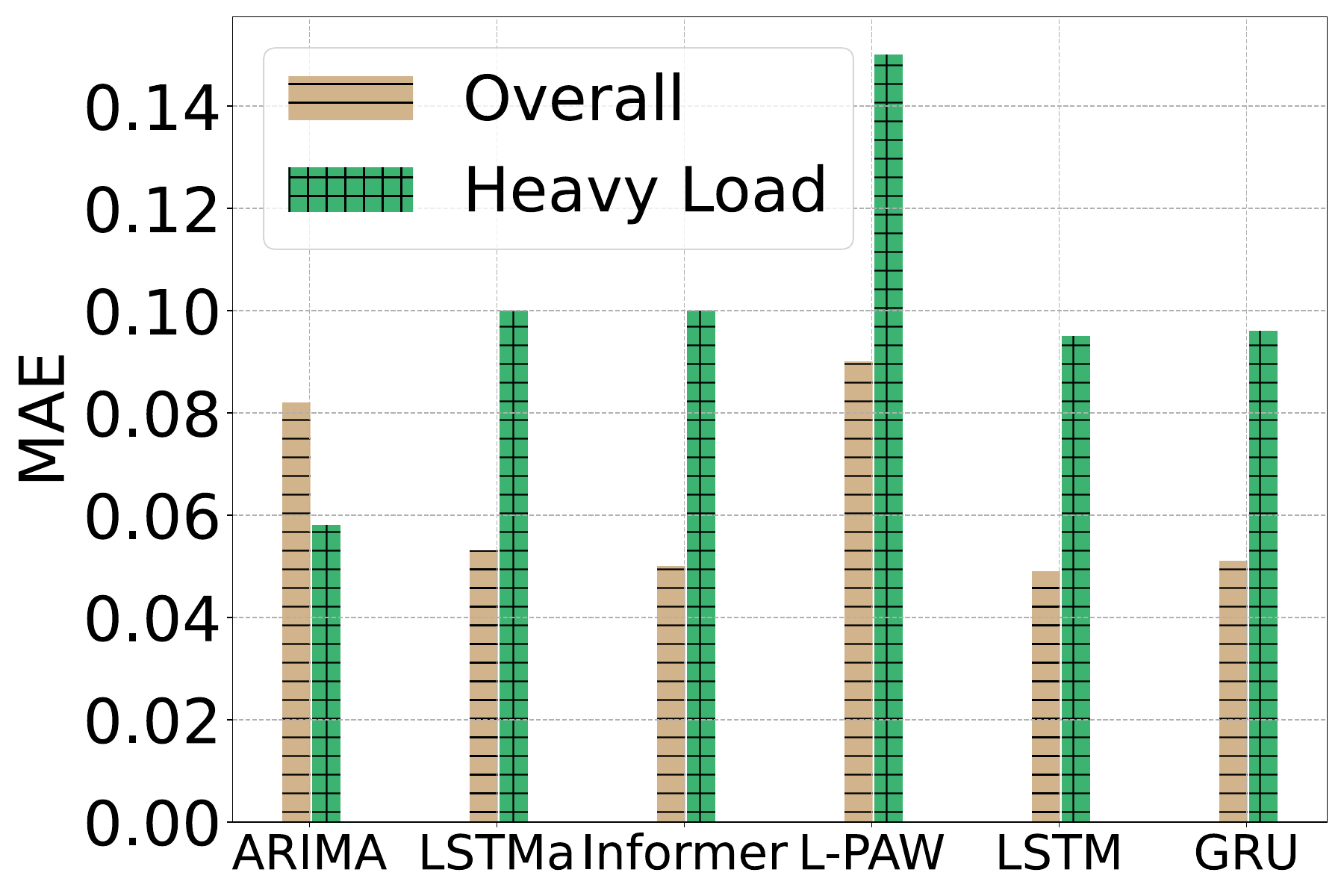}
  \label{fig:Accuracy}
  }
  \hfill
  \subfigure[Prediction of GRU on Alibaba2018.]{
  \includegraphics[width=0.45\linewidth]{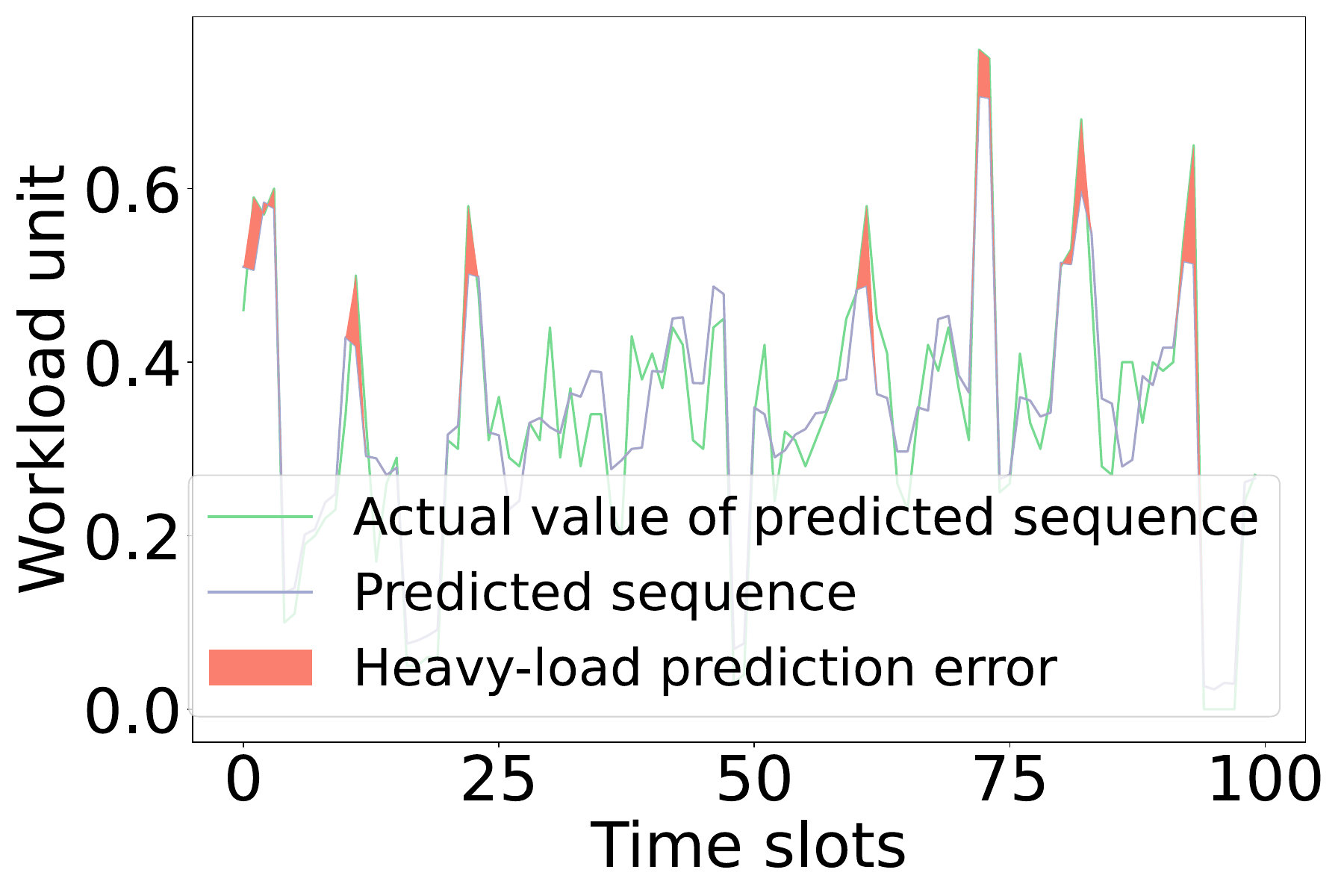}
  \label{fig:peakPred}
  }
  \hfill
  \caption{(a)-(b) Overall workload and heavy workload prediction accuracy}
\end{figure}

However, the high variability of cloud-server workload \cite{chen2019towards,10.1145/3308560.3316466} makes workload prediction challenging. For time-series forecasting methods that rely on periodicity information \cite{guo2019attention,chen2019gated,lv2018lc,yao2019revisiting}, they usually assume fixed and known periodic length, which is not in line with dynamic cloud environments where periodic patterns vary across services and over time. Those SOTA methods that do not highly rely on periodicity \cite{wu2022timesnet,das2024decoder,jin2023time,goswami2024moment,zhou2023one,shi2024time,DBLP:conf/icml/LiuZLH0L24} show better generalization performance than those do so. However, they are vulnerable to data imbalance between heavy workload and common one and have been reported much lower performances on rare heavy workload burst \cite{9178492,10.1145/3292500.3330896}.
As an empirical example, Fig. \ref{fig:Accuracy}- \ref{fig:peakPred} show the overall prediction accuracy and heavy-workload one of some prevailing methods (ARIMA \cite{calheiros2014workload}, LSTM with attention (LSTMa) \cite{zhu2019novel}, Informer \cite{zhou2021informer}, Learning based Prediction Algorithm for cloud workload (L-PAW) \cite{chen2019towards}, LSTM and GRU), where the latter's error is nearly twice as large as the former's error. 
However, accurately predicting heavy workloads is more important than predicting regular ones, because errors in heavy-workload prediction are more likely to cause SLA violations than those in normal-load prediction, as they provide misleading information to the scheduler during workload bursts.

To cope with time series with changeable periodicity and to accurately predict heavy workloads, we propose PePNet, which enables accurate prediction of highly variable workloads, including rarely occurring heavy-workload events.
It improves the prediction accuracy in two aspects: 1) mining and fusing periodic information adaptively when periodicity is changeable without any priori knowledge; and 2) using an \emph{Achilles' Heel Loss Function} (AHLF)~\footnote{Achilles' Heel originates from Greek mythology and is the fatal weakness of the hero Achilles}, which pays more attention to the under-fitting part to offset the negative influence of data imbalance.
To address the problem of changeable periodicity in cloud environment, we propose a Periodicity-Perceived Mechanism with a Periodicity-Mining Module for automatic periodicity and period-length detection and a Periodicity-Fusing Module for adaptively handling mixed periodicity. We further provide a theoretical error bound for the mined periodic information and an automatic hyperparameter selection strategy.
AHLF focuses on the upper bound of prediction error by iteratively minimizing the most under-fitted segment in a prediction sequence. This alternating fitting strategy mitigates the overemphasis on sporadic patterns observed in many loss functions combatting the under-fitting problem \cite{ding2019modeling}, thereby alleviating under-fitting under heavy load and improving heavy-load and overall prediction accuracy.

We aim to make the following novel contributions to the field of workload prediction in cloud computing:
\begin{itemize}
  \item  We propose a Periodicity-Perceived Mechanism that mines and adaptively fuses periodic information for time series with changeable periodicity, with a theoretical error bound and an automatic hyperparameter selection strategy.
  \item  We formulate an AHLF to improve heavy-workload and overall prediction accuracy by mitigating the overemphasis on sporadic patterns.
  \item We perform extensive experiments on Alibaba2018, Server MAchine Dataset (SMD), and Dinda's datasets to show that PePNet improves heavy-workload Mean Square Error (MSE) by 21.0\% and overall MSE by 11.8\% on average.
\end{itemize}

\begin{figure*}[tbhp]
  \centering 
  \includegraphics[width=\linewidth]{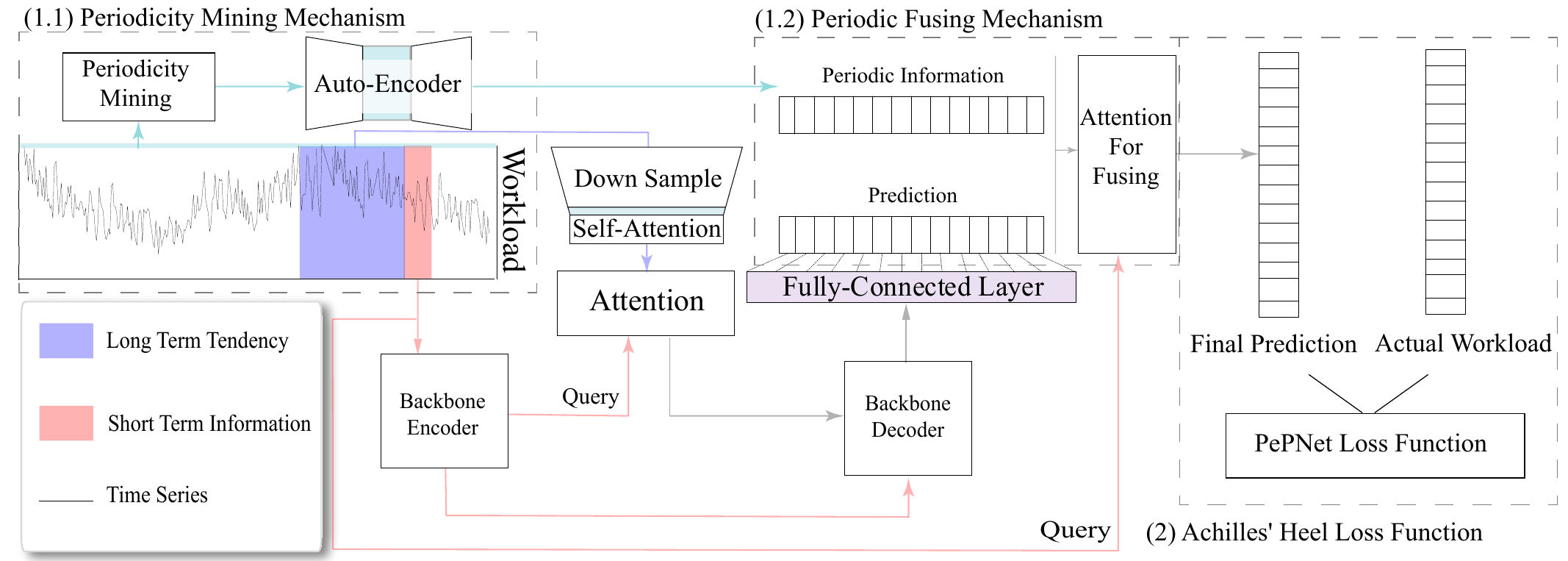}
  \caption{The overview of PePNet.}
  \label{modelArch}

\end{figure*}

\section{Methodology}
\textbf{Preliminary}.
Input data $X$ is divided into three parts, which are denoted by $\bar{X} \in \mathbb{R}^{I \times d}$, $\tilde{X} \in \mathbb{R}^{M \times d}$, $X_{p} \in \mathbb{R}^{P \times d}$ respectively. The $d$ is the dimension of a feature vector at each time slot and $I, M$ and $P$ respectively stand for the length of short-term dependent information, long-term tendency information and periodic information. 
$\bar{X}$ is the nearest workload time series before the predicting part and shows high relevance to predicting workload;
$\tilde{X}$ is a bit longer workload time series before $\bar{X}$ and reflects the long term tendency in workload variation;
$X_{p}$ is the first period of workload for each machine; $Y$ is the ground truth value of predicted workload. 
We use $\tilde{X}_i$ and $\bar{X}_i$ to denote the $i$-th time slot's workload in $\tilde{X}$ and $\bar{X}$ respectively.

\subsection{Overview of PePNet}
PePNet is based on an encoder-decoder \cite{cho2014learning} architecture, which can be implemented as a transformer, LSTM, etc. It fuses three kinds of information from rough data: long-term tendency, the short-term dependent and periodic information. Besides, the whole model is trained by minimizing AHLF to offset the negative impact of data imbalance. The overview of PePNet is shown in Fig. \ref{modelArch}.

In the encoder, the processes of extracting short-term, long-term and periodic information are shown as follows. In function $\operatorname{Attention}(q,k,v)$, $q$, $k$, and $v$ respectively stand for the query, key and value\cite{NIPS2017_3f5ee243}. PeriodicityMining module is introduced in subsection~\ref{sec:ppmech}. The downsample module downsamples the input time series using an increasing arithmetic sequence as the stride.
\begin{equation}
    \dot{\bar{X}}_{i+1} = \operatorname{Encoder}(\bar{X}_i)
    \label{enc_lstm}
\end{equation}
\begin{equation}
    y_{p},X_{p}=\operatorname{PeriodicityMining}(X,\bar{X})
\end{equation}
\begin{equation}
    \dot{\tilde{X}}= \operatorname{Downsample}(\operatorname{Attention}(\tilde{X},\tilde{X},\tilde{X}))
    \label{enc_att}
\end{equation}

The process of decoder is given as follows, where $W_h, W_c$ are both model parameters and the PreiodicityFusing module is introduced in subsection~\ref{sec:ppmech}.
\begin{equation}
    \hat{y}_{j+1} = \operatorname{Decoder}(\hat{y}_j,h)
    \label{dec_lstm}
\end{equation}
\begin{equation}
    h=\operatorname{Attention}(\bar{X},\dot{\tilde{X}},\dot{\tilde{X}})
    \label{dec_h}
\end{equation}

\begin{equation}
    y=\operatorname{PeriodicityFusing}(\bar{X},y_{p},\hat{y})
    \label{dec_fuse}
\end{equation}

\subsection{Periodicity-Perceived Mechanism}
\label{sec:ppmech}
Periodic information can be used to effectively improve the overall prediction accuracy as well as heavy workload. It consists of \emph{Periodicity-Mining Module} and \emph{Periodicity-Fusing Module}.

%There are three main challenges to fuse periodic information. 
%(1) We have no priori knowledge about the periodicity of workload for different machines, which calls for the \emph{Periodicity-Mining Module} to detect the existence of periodicity and the length of one period. 
%(2) The periodicity for different machines is variable (i.e., strict periodic, lax periodic, aperiodic), which calls for an adaptive \emph{Periodicity-Fusing Module}. 
%(3) It is hard to fuse the periodic information in lax periodic series because lax periodic series are sometimes periodic and sometimes not. In Fig. \ref{fig:laxP}, we overlap a series shifted one period ahead and the original series. As Fig. \ref{fig:laxP} shows, there are mainly two obstacles to fuse periodic information in lax periodic series: periodic shift and local periodicity violation. The former one can be solved by dynamic matching which is depicted in section \emph{Periodicity-Mining Module}. The latter is caused by noise and external events. Therefore, we filter out the noise in periodic information in \emph{Periodicity-Mining Module} and evaluate the reliability of periodic information in the \emph{Periodicity-Fusing Module}.

\subsubsection{Periodicity-Mining Module}
\label{PerMine}
\begin{figure*}[t]
    \centering
    \subfigure[Autocorrelation coefficient for periodic sequence.]{
    \includegraphics[width=0.2\linewidth]{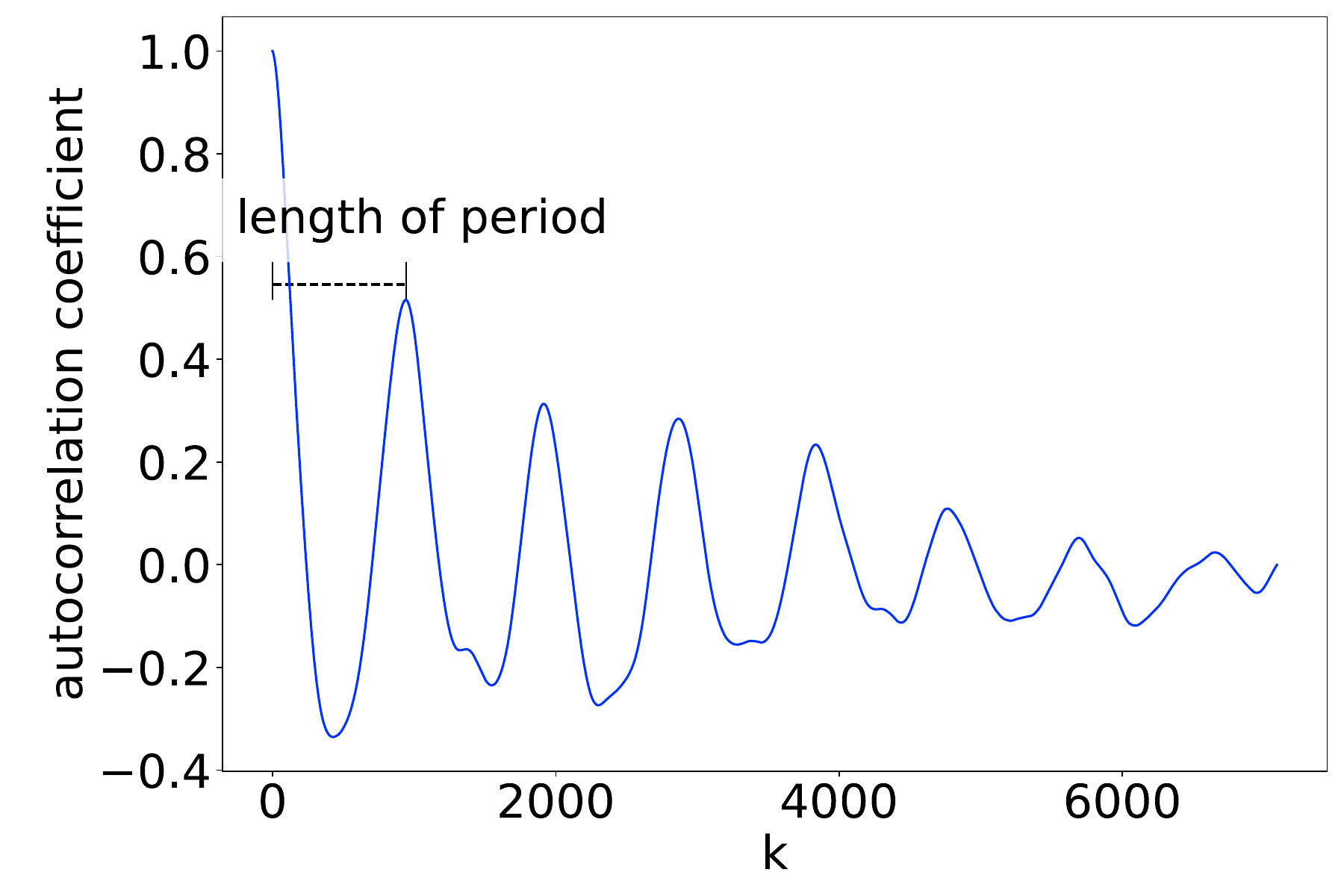}
    \label{fig:auto}
    }
    \hfill
    \subfigure[Autocorrelation coefficient for aperiodic sequence.]{
    \includegraphics[width=0.2\linewidth]{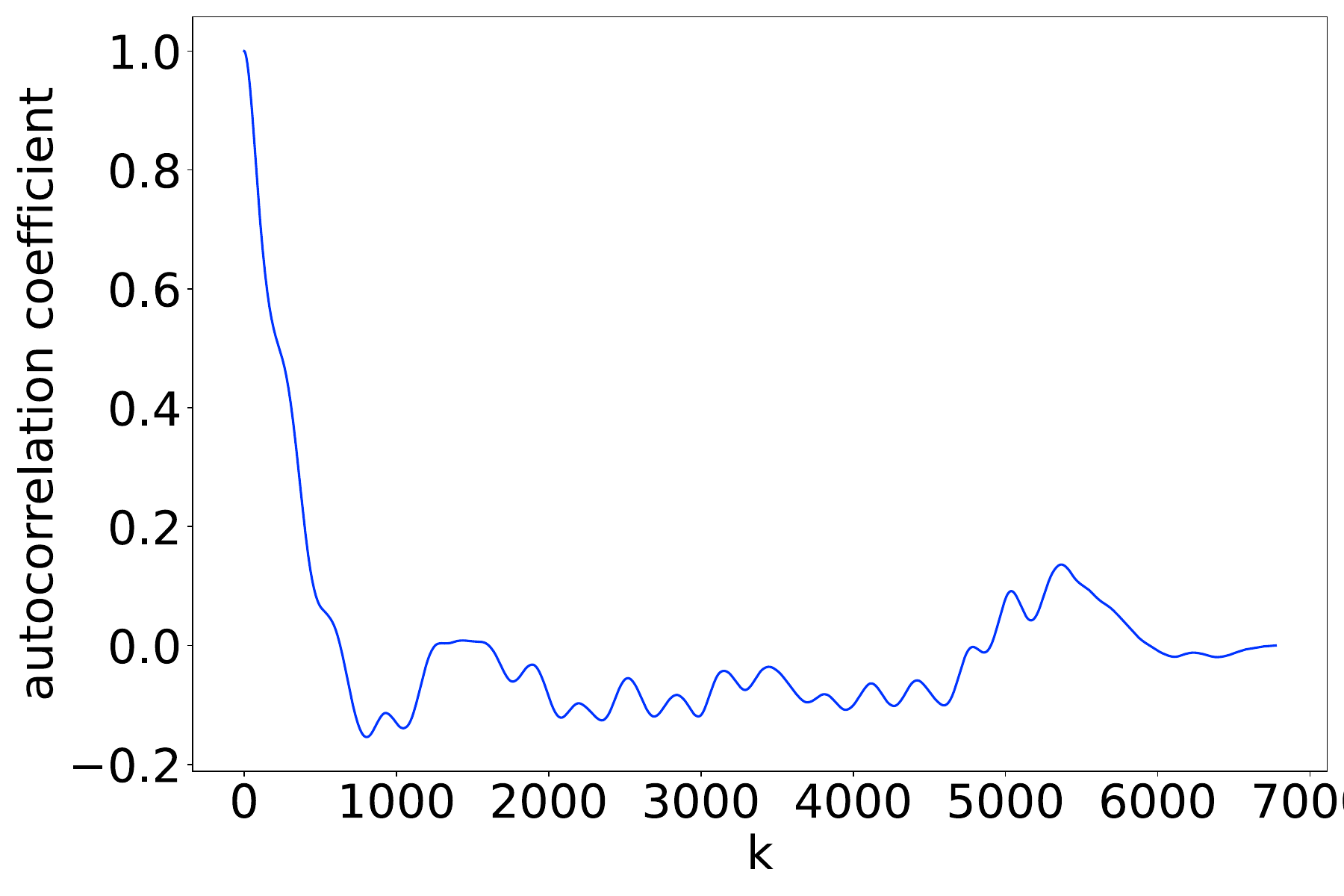}
    \label{fig:npAuto}
    }
    \hfill
    \subfigure[Histogram of values for the first peak of the autocorrelation coefficient.]{
        \includegraphics[width=0.2\linewidth]{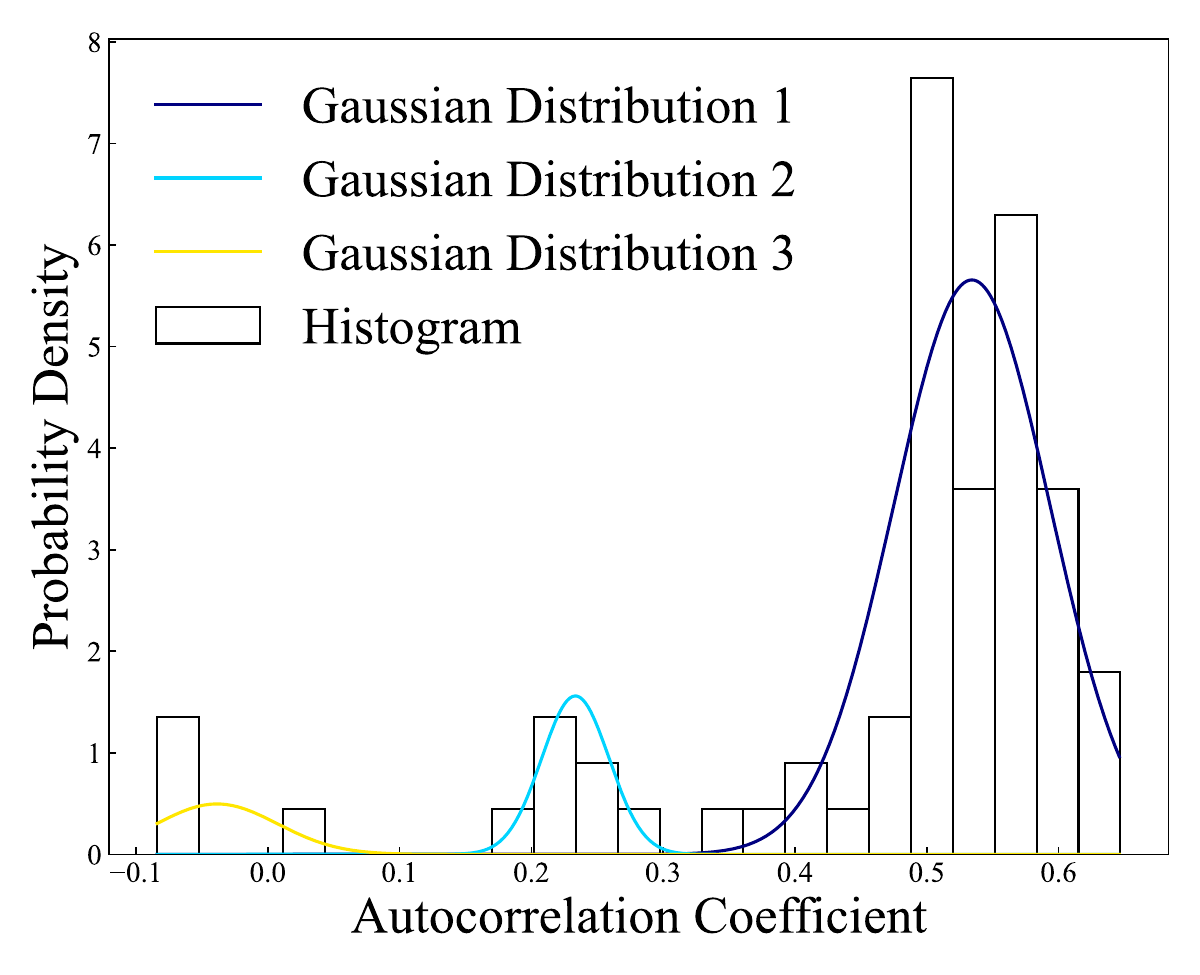}
        \label{histo}
    }
    \hfill
    \subfigure[How the value of $\gamma$ impacts the distribution of gradient weights.]{
        \includegraphics[width=0.2\linewidth]{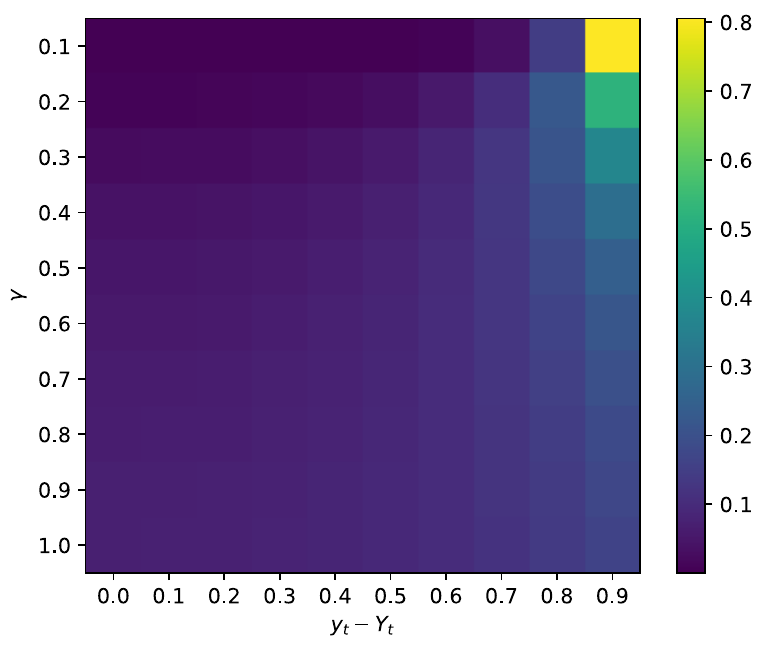}
        \label{gammaEffect}
    }
    \hfill
    \caption{Illustrations for data division and periodicity-perceived mechanism.}

\end{figure*}

PePNet calculates the time series autocorrelation coefficient $\rho_k$ for each machine as:
\begin{equation}
    \rho_k=\frac{\operatorname{cov}(X_{t-k},X_{t})}{\sqrt{\operatorname{cov}(X_{t},X_{t})\operatorname{cov}(X_{t-k},X_{t-k})} }
    \label{autoCorr}
\end{equation}
which represents the linear correlation among all workloads with interval $k$ (i.e., linear correlation between $X_{t}$ and $X_{t-k}$, for every $t\in \{t\in \mathbb{N}^+|t<l\}$, $X_{t}$ stands for the workload at the $t$-th time slot). As shown in Fig. \ref{fig:auto}, when the workload is periodic, the autocorrelation coefficient rises to a large value again after the first decline, which represents a high hop relevance of workload. While the sequence is aperiodic, the autocorrelation coefficients are shown in Fig. \ref{fig:npAuto}, which have a distinct pattern. Therefore, PePNet sets a hyperparameter $\mathcal{T}$ and judges whether a time series is periodic by detecting whether the autocorrelation coefficient crosses over $\mathcal{T}$ again. The value of $\mathcal{T}$ denotes the acceptable range of periodicity strictness.

The position of the first peak in autocorrelation coefficients of periodic sequence strikes the highest hop relevance and denotes the length of one period. We illustrate it in Fig. \ref{fig:auto}. The reason why the autocorrelation coefficient shows such a pattern is that the user behavior is continuous in time, thus autocorrelation coefficients of small $k$ are large. As $k$ increases, the relevance drops down. But when $k$ reaches around the integer multiple of period length, the workload at the two moments $X_t$ and $X_{t-k}$ shows a high linear correlation again.  

Based on these observations, PePNet traverses the value of $k$ from small to large. If PePNet finds the smallest $k$ that makes the autocorrelation coefficient $\rho_k$ satisfy the conditions: $\rho_k>\rho_{k-1}, \rho_k>\rho_{k+1}$ and $\rho_k> \mathcal{T}$, the time series is periodic and the length of one period is $k$. Otherwise, it is aperiodic. We cut out the first period of the machines whose workloads are periodic in the training set as a periodic information knowledge base, which is denoted by $X_{p}$. To solve the issue of period shift, PePNet uses a dynamic matching. If the workload is periodic, PePNet finds the sequence $\{X_{t_a},\ldots,X_{t_a+I-1}\}$ in $X_{p}$, whose distance from $\bar{X}$ is the smallest. The distance can be computed either by MSE or by Dynamic Time Warping (DTW). Then, PePNet sets $y_{p}=\{X_{t_a+I},\ldots,X_{t_a+I+J}\}$. Otherwise, $\{-1,-1,\ldots,-1\}$ is set to $y_{p}$. 

\emph{Error estimation.} We use $E[(X_{t-k}-X_t)^2]$ to measure the quality of periodic length $k$ found by \emph{Periodicity-Mining Module}. We have the following result, where $\sigma_{t-k}, \sigma_{t} , \mu_{t-k},$ and $ \mu_{t}$ are the standard deviation and expectation at $t-k$ and $t$ time slots respectively, $\Delta\sigma=\sigma_{t-k}-\sigma_{t}$ and $\Delta\mu=\mu_{t-k}-\mu_t$. $\sigma$ is the standard deviation of a stationary time series.

\textbf{Theorem 1.} When using \emph{Periodicity-Mining Module}, $E[(X_{t-k}-X_t)^2]$ is upper-bounded by $(\Delta\sigma)^2+(\Delta\mu)^2+2(1-\mathcal{T})\sigma_{t-k}\sigma_{t}$. Furthermore, when the time series is stationary, $E[(X_{t-k}-X_t)^2]$ is upper-bounded by $2(1-\mathcal{T})\sigma^2$. 

Proof of Theorem 1 is given in Appendix~\ref{app:theo1}.

Taking a further look at the upper bound, its first and second items are constants. Thus, the value of error depends on the last item, which is determined by the value of $\mathcal{T}$. Considering that $\mathcal{T}$ is less than or equal to one, the closer $\mathcal{T}$ is to one, the smaller the error is. However, most of the time series are lax-periodic in reality. Setting $\mathcal{T}$ as 1 may miss many lax-periodic time series. Thus, there is a tradeoff between improving the accuracy of periodic length and finding all the periodic time series (including lax-periodic ones). 

Considering that tradeoff may make it challenging to choose a proper value of $\mathcal{T}$, we provide an automatic setting method for $\mathcal{T}$ based on statistical observations. For finer tuning, this method provides a searching range for autocorrelation coefficient first peak threshold $\mathcal{H}$.
Taking Alibaba2018-CPU\footnote{https://github.com/alibaba/clusterdata} as an example, we take 75 machines in it and plot a histogram of first-peak value in Fig. \ref{histo}. There are two properties of the distribution: clustered and following Gaussian Distribution for each cluster. We use Gaussian Mixture Model (GMM) \cite{wu2020self} to fit the mixed Gaussian distribution of its first peak values and plot the Gaussian distributed curve on the histogram in Fig. \ref{histo}. Different clusters have different degrees of periodicity. The higher the expectation of the cluster is, the stricter its periodicity is. To guarantee the quality of the periodicity that \emph{Periodicity-Mining Module} detects, it picks the workload in the cluster with the highest expected value as periodic and filters the others.
Let $\mu$ and $\sigma$ denote the expectation and standard deviation of the cluster with the highest expectation. In industrial applications, $\mu-\sigma$ can be directly used as $\mathcal{H}$ to save the labor cost, as the probability that the first peak is greater than $\mu-\sigma$ in this cluster is 84.2\% according to Gaussian distribution. For finer tuning, $\mathcal{H}$ can be explored between $\mu-\sigma$ and $\mu+\sigma$.

\subsubsection{Periodicity-Fusing Module}
To filter out random noise in the periodic information, PePNet firstly uses an auto-encoder \cite{10.1007/978-3-642-21735-7_7}.
To solve the problem of variable periodicity for different machines and the problem of local periodicity violation, PePNet uses an attention mechanism to evaluate the reliability of periodic information. The final prediction $y$ is given as:
\begin{equation}
    y_{p}^{'}= \operatorname{Autoencoder}(y_{p})
    \label{filtery}
\end{equation}
\begin{equation}
    y=\operatorname{Attention}(\bar{X},(\hat{y},y_{p}^{'}),(\hat{y},y_{p}^{'}))
    \label{y_equ}
\end{equation}

\subsection{Achilles' Heel Loss Function (AHLF)}
\label{Sec:lossf}
Heavy loads occur infrequently, causing severe data imbalance and poor prediction performance. To address this issue, PePNet adopts the AHLF, which focus on the part with largest prediction error. During training, it prioritizes the most under-fitted regions. As these errors are reduced, the focus adaptively shifts to other under-fitted parts. This iterative process effectively improves heavy-load prediction accuracy while maintaining high overall forecasting performance.
AHLF is given as:
\begin{equation}
    \label{lossT}
    l(Y,y)=\gamma \log(\sum_{t\in T} exp(\frac{(y_t-Y_t)^2}{\gamma})), \gamma>0
\end{equation}
where $T=\{t_1,\ldots,t_{1+J}\}$, which are the time slots of a forecasting sequence. $y_t$ denotes the predicted workload at time $t$ (the prediction length for one sample should be at least 2), $Y_t$ denotes the ground-truth workload at time $t$, and $\gamma$ is a scale factor.

AHLF actually distributes more weight to the gradient for higher prediction error when back propagating. This aspect is obvious by taking first deriving of Eq.~(\ref{lossT}), i.e:
\begin{equation}
    \label{gradient}
    \frac{\partial l(Y,y)}{\partial \mathcal{P}}
    =\sum_{t=1}^T(\alpha_t \cdot \frac{\partial{(y_t(\mathcal{P})-Y_t)^2}}{\partial \mathcal{P}})
\end{equation}
where $\alpha_t=\frac{exp{\frac{(y_t(\mathcal{P})-Y_t)^2}{\gamma}}}{\sum_{t=1}^T \exp{\frac{(y_t(\mathcal{P})-Y_t)^2}{\gamma}}}$, $\mathcal{P}$ stands for the parameters of PePNet. $\alpha_t$ is actually a normalized weight whose value depends on the prediction error at time slot $t$. $\gamma$ is a scale factor. When $\gamma$ is large, the difference in prediction errors across time slots is narrowed and the weights are relatively uniform for each time slot. When $\gamma$ is small, the difference in prediction errors across time slots is amplified and the weights become polarized. To visually demonstrate this point, we plot the value of $\alpha_t$ for different combinations of $\gamma$ and prediction error $y_t-Y_t$ in Fig. \ref{gammaEffect}. In it, each row shows the value of $\alpha_t$ for a set of prediction error $\{y_t-Y_t|0 \leq t \leq 9\}$, when specifying $\gamma$. 
For a specific $\gamma$, the larger the prediction error is, the bigger $\alpha_t$ is. Besides, for a specific prediction error in a specific error set, the smaller $\gamma$ is, the bigger $\alpha_t$ is. When $\gamma$ is set to 0.1, $\alpha_t$ for the largest prediction error in this row approximates 1.

It is worth noticing that when $\gamma$ is indefinitely close to 0, AHLF becomes a simple formation, as given in the following theorem. 

\textbf{Theorem 2.} As $\gamma$ is infinitely close to 0, the gradient of the loss function infinitely approximates: 
\begin{equation}
    \label{grad0}
    \frac{\partial g(t^*,\mathcal{P})}{\partial \mathcal{P}}
\end{equation}
where $t^* = \mathop{\arg\max}_{t} (y_t-Y_t)^2$ and $g(t^*,\mathcal{P}) = (y[t^*]-Y[t^*])^2$.

Proof of Theorem 2 is given in Appendix.~\ref{app:theo2}.

The theorem implies that when $\gamma$ indefinitely approximates 0, only the gradient for maximum prediction error is assigned a weight of 1, while others are assigned a weight of 0.

\section{Experiment}
In this section, we conduct extensive experiments to validate the following findings:
1) PePNet improves the accuracy of heavy-workload prediction as well as the overall one compared with the state-of-the-art methods;
2) PePNet only introduces slight time overhead for training and inference over the existing methods;
3) PePNet is insensitive to hyperparameters and has high robustness;
4) Each mechanism in PePNet is proven to play an important role via ablation studies; and
5) Due to its adaptability for periodicity, when the time series is aperiodic, PePNet still works well.

\begin{table}[hbtp]
    \centering
    \renewcommand\arraystretch{1}
    \caption{\label{Parameter}Basic hyperparameters of PePNet.}
    %\setlength\tabcolsep{3pt}
    %\resizebox{\linewidth}{!}{%
    \begin{tabular}{c|c}
        \hline
        \textbf{Hyperparameter}         & \textbf{Value} \\ \hline
        Batchsize of Alibaba2018-CPU        & 100            \\ 
        Learning rate of Alibaba2018       & 0.001          \\ 
        Learnig rate of Dinda's dataset & 0.0005          \\ 
        Learning rate of SMD            & 0.01 \\ 
        $\mathcal{T}$ of Alibaba2018-Mem    & 0.2  \\ 
        $\mathcal{T}$ of others             & 0.5  \\ 
        $\gamma$                        & $\mathop{\lim}_{\gamma\rightarrow 0}$ \\ 
        \hline
        \end{tabular}%

\end{table}

\subsection{Experiment Setup}

\textbf{Hyperparameters.} We summarize the most important hyperparameters of PePNet in Tab. \ref{Parameter}. 

\textbf{Baseline Methods.} We compare PePNet with the state-of-the-art time series predicting methods and popular workload predicting models and the variants of PePNet : Autoformer (Autof) \cite{NEURIPS2021_bcc0d400}, LSTM with attention mechanism (LSTMa) \cite{zhu2019novel}, Informer (Inf) \cite{zhou2021informer}, L-PAW \cite{chen2019towards}, LSTM \cite{6795963}, GRU \cite{cho2014learning}, Reformer (Ref) \cite{DBLP:conf/iclr/KitaevKL20}, Fedformer (Fedf) \cite{DBLP:conf/icml/ZhouMWW0022}, and Time-MoE (TMoE) \cite{shi2024time}. Besides, we implement the backbone-encoder and backbone-decoder as transformer and LSTM respectively and denote them as PeP-T and PePNet. Moreover, we also introduce some variants for ablation study. We use PeP$^-$ to denote PePNet that removes the Periodicity-Perceived Mechanism. We use PeP$^\dagger$ to denote the PePNet that uses MSE as a loss function. We use PeP$^\ddagger$ to denote the PePNet that both removes the Periodicity-Perceived Mechanism and does not use AHLF but uses MSE.

\begin{figure*}[htbp]
    \centering
    \subfigure[Data distribution of different datasets]{
    \includegraphics[width=0.2\linewidth]{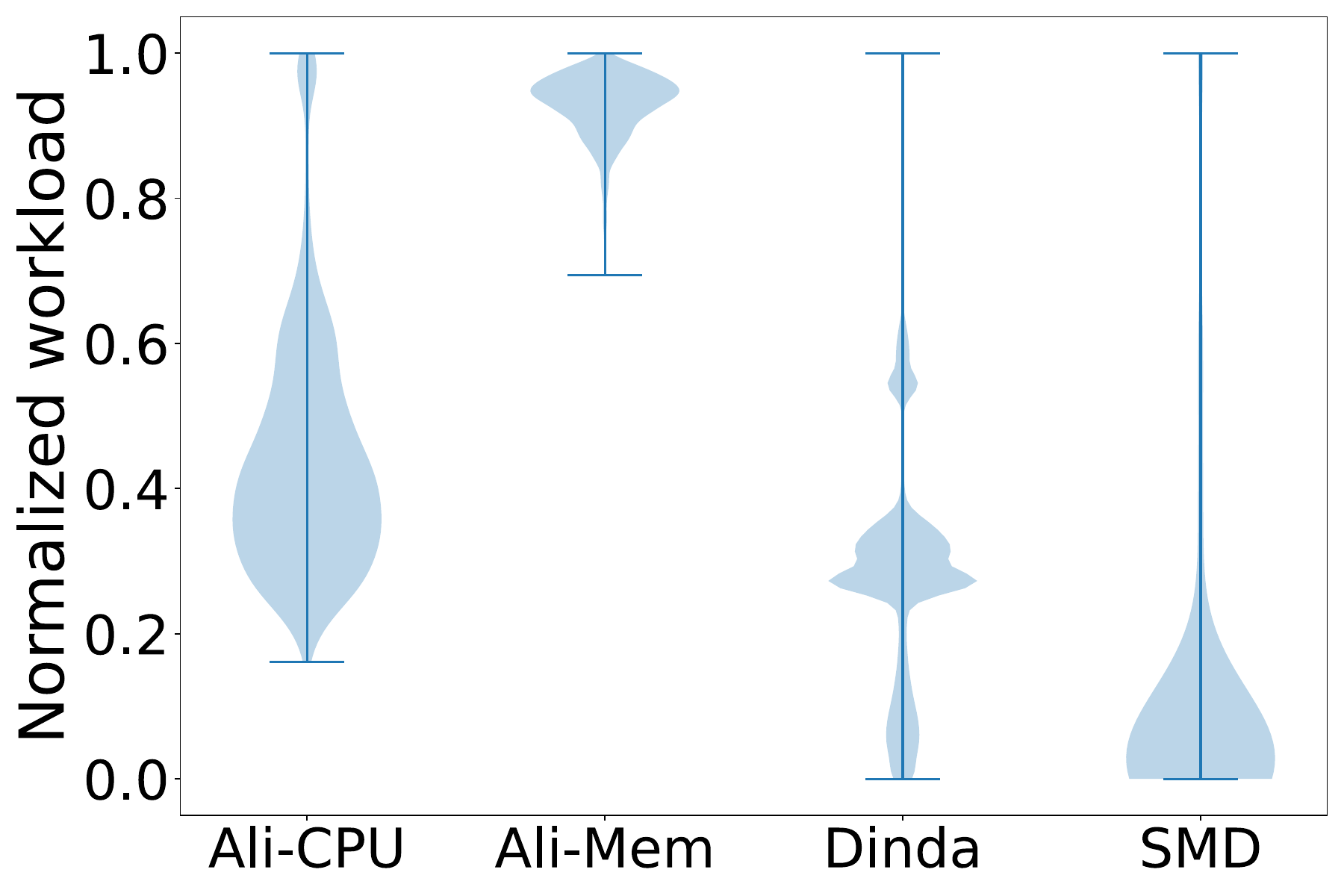}
    \label{fig:dataDis}
    }
    \hfill
    \subfigure[Time overhead of different methods]{
    \includegraphics[width=0.2\linewidth]{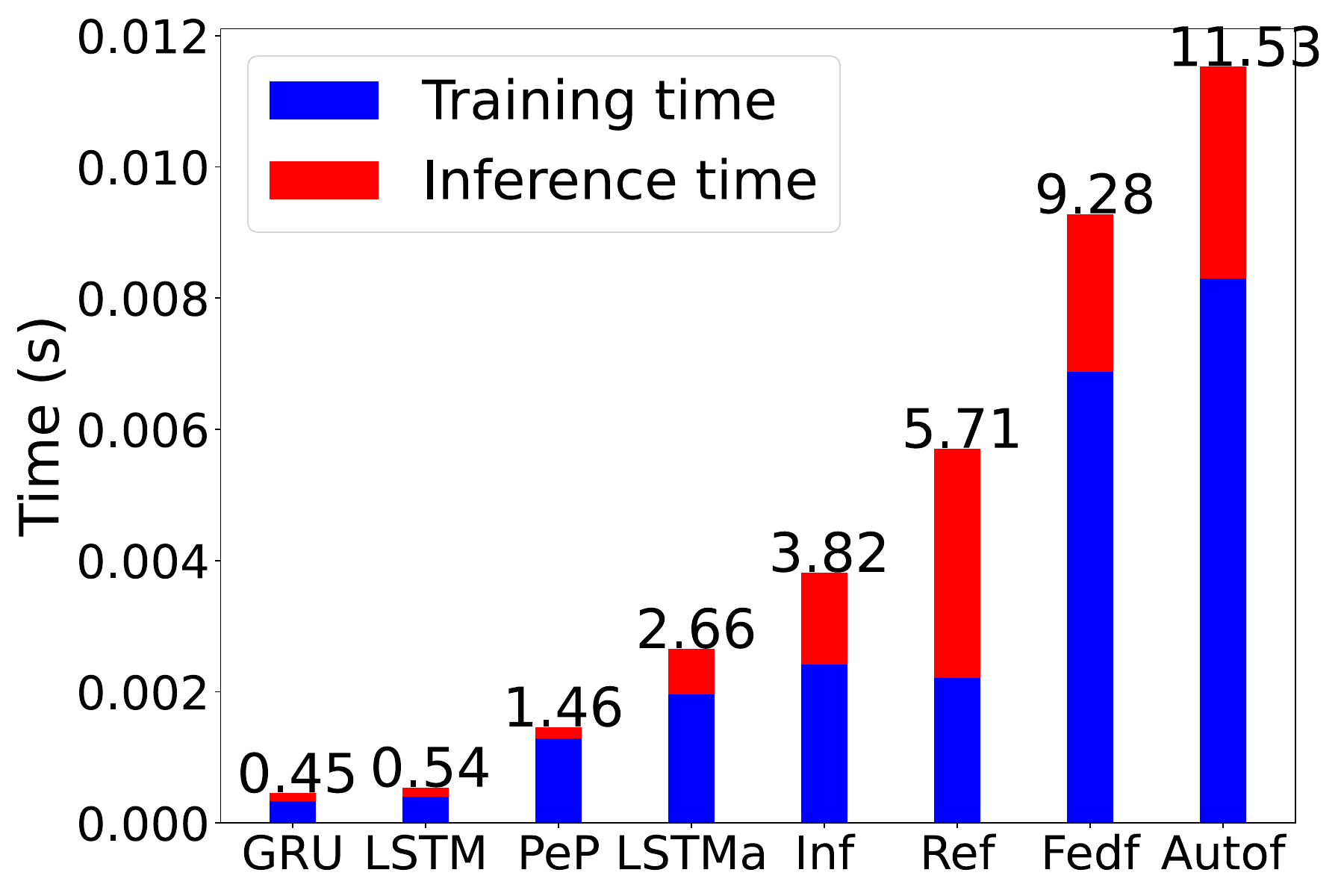}
    \label{fig:Time}
    }
    \hfill
    \subfigure[MAE of PePNet with different $\gamma$]{
    \includegraphics[width=0.2\linewidth]{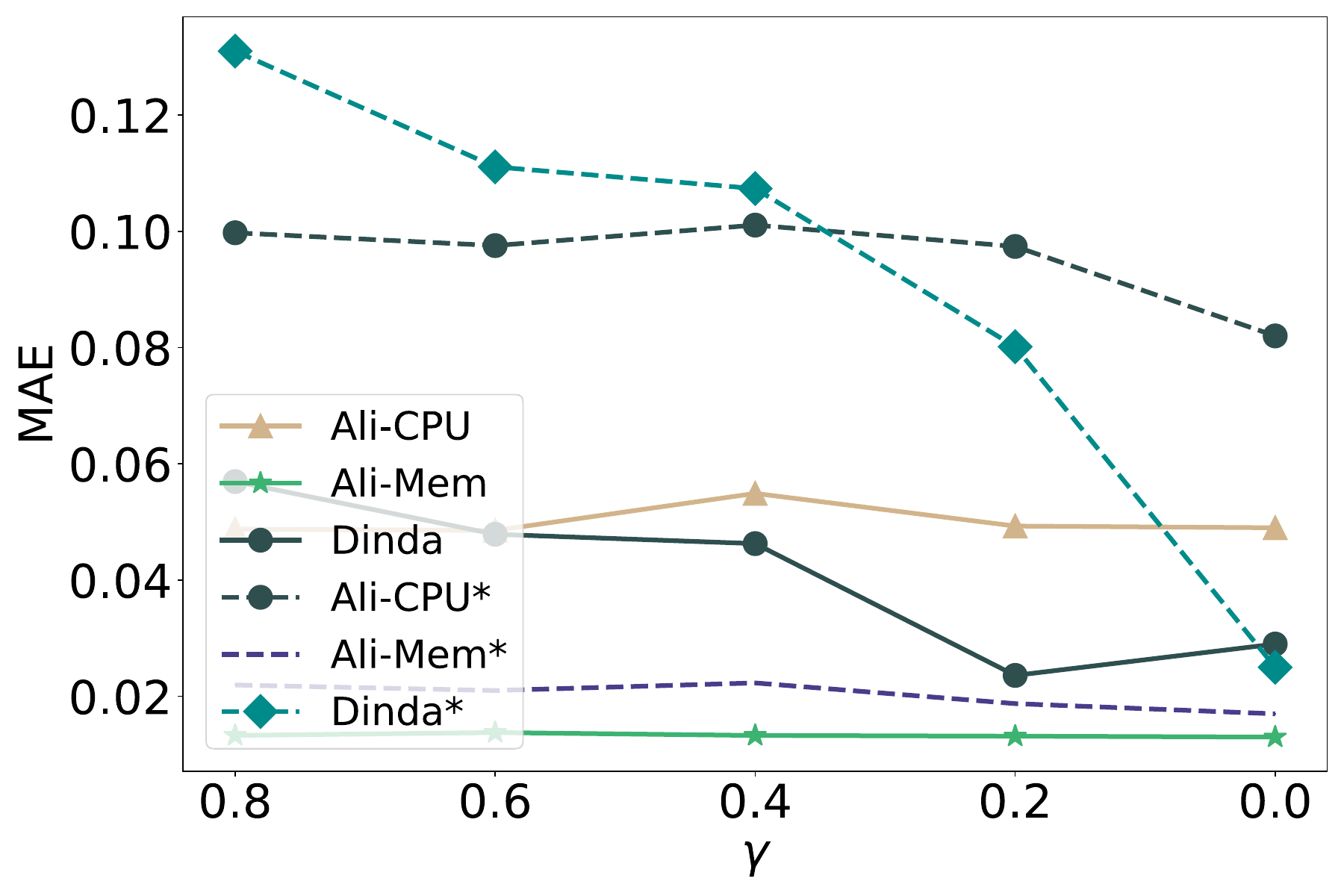}
    \label{fig:Gamma}
    }
    \hfill
    \subfigure[Overall MAE of PePNet on Alibaba2018-CPU]{
    \includegraphics[width=0.2\linewidth]{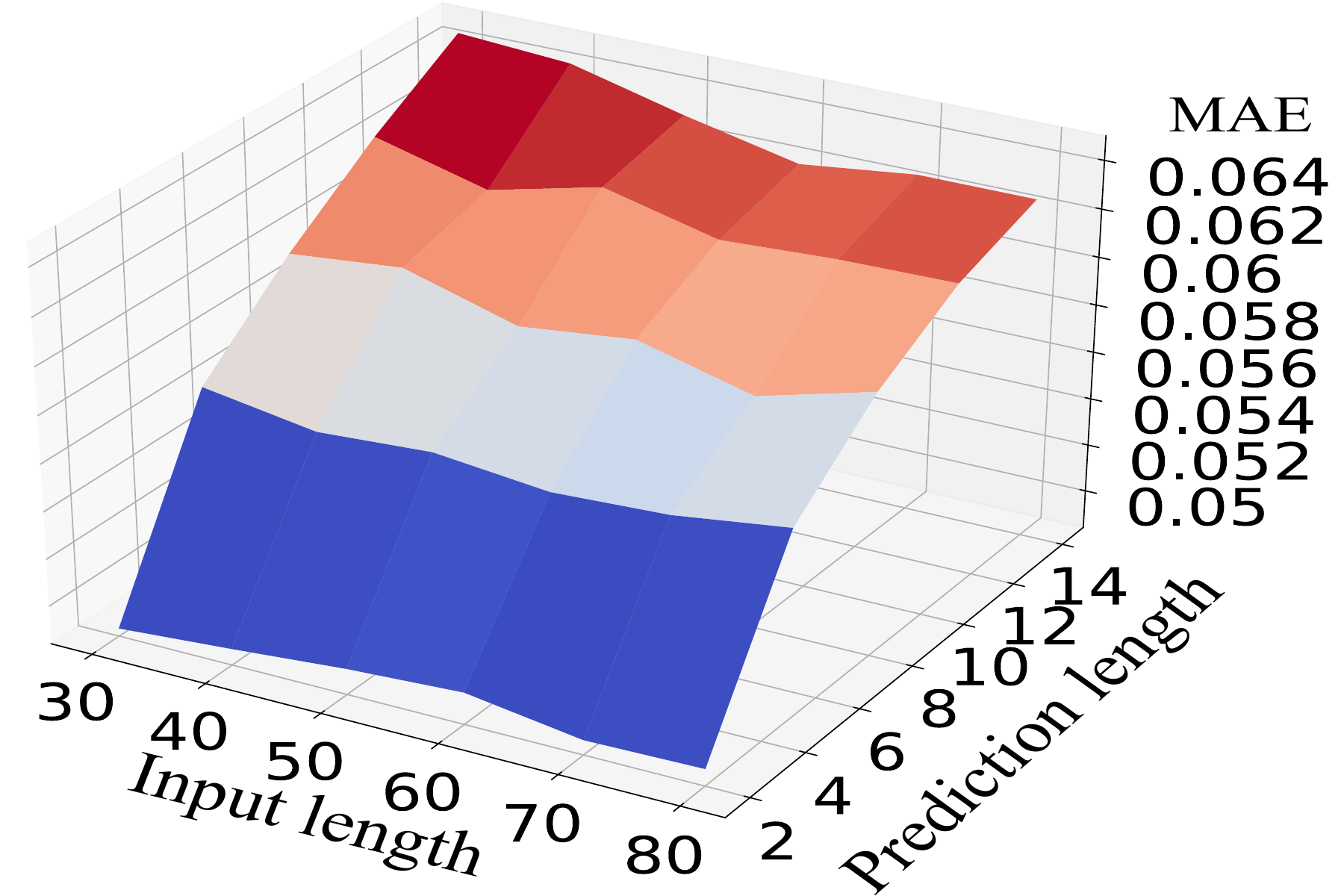}
    \label{fig:cpuO}
    }

    \caption{(a) The value of the y-axis is the original workload divided by the maximum workload of its machine. The wider the width of the violin in the figure is, the more workloads are distributed near this value. (b) The figure shows the average consuming time of one forward propagation and backpropagation for one sample. The labels on the histogram are 1000 times of the actual value. Besides, we use the shorthand of Informer, Autoformer and Reformer in the xticks. (c) The figure shows the overall and heavy-workload MAE of PePNet for different $\gamma$ on all the three datasets. In the legend, we use dataset name* to represent the heavy-workload MAE on the specific dataset, while using dataset name to represent the overall MAE on the dataset. (d) The figure show overall MAE of PePNet with different combination of the length of $X_{in}$ and $p$.}

\end{figure*}

\textbf{Datasets.} We perform experiments on three datasets: Alibaba's cluster trace v2018\footnote{https://github.com/alibaba/clusterdata}, Dinda's dataset collected from Pittsburgh Supercomputing Center \cite{dinda1999statistical} and Server Machine Dataset (SMD) \cite{su2019robust} collected from one of global top-10 Internet companies. From different organizations with different system configurations, they can help us to verify the generalization of PePNet.
We plot the data distribution of different datasets in Fig. \ref{fig:dataDis}. It is worth noting that different datasets represent different data distributions: long-tailed distribution (Dinda's dataset), centralized distribution (memory usage of Alibaba2018 and SMD), and relatively even data distribution (CPU usage of Alibaba2018). 
We use the first 1,000,000 lines in Alibaba's dataset and \emph{axp0} in Dinda's dataset. The values in Alibaba2018 are utilization percentage and we uniformly scale it to range [0, 1].

\textbf{Evaluation metrics.} We use three metrics to evaluate performance of all the methods: MSE, Mean Absolute Error (MAE), and Mean Absolute Percentage Error (MAPE).

\begin{table*}[t]
\centering
\renewcommand\arraystretch{1.1}
\setlength\tabcolsep{1pt}

\caption{The overall and heavy-workload prediction accuracy on three datasets.
The MSE in the table are multiplied by 1000.}
\label{Result}

\resizebox{\textwidth}{!}{%
\begin{tabular}{c|c|c|ccccccc|ccccc}
        \hline
                &            &      & \multicolumn{1}{c}{LSTMa} & \multicolumn{1}{c}{Inf} & \multicolumn{1}{c}{L-PAW} & \multicolumn{1}{c}{Ref} & \multicolumn{1}{c}{Autof} & \multicolumn{1}{c}{Fedf}  & \multicolumn{1}{c|}{TMoE}  & \multicolumn{1}{c}{PeP$^-$} & \multicolumn{1}{c}{PeP$^\dagger$} & \multicolumn{1}{c}{PeP$^\ddagger$} & \multicolumn{1}{c}{PeP-T} & \multicolumn{1}{c}{PePNet}\\ \hline
                &            & MAPE & 0.117                     & \underline{0.100}       & 0.102                     & 0.104                   & 0.211                     & 0.110                     &  0.752                         & 0.089                 & 0.087                 & 0.116                  &   \textbf{0.047}            & 0.076              \\
                & Overall    & MSE  & 8.233                     & \underline{4.865}       & 10.960                    & 8.950                   & 37.636                    & 9.211                     &  5.171                         & 5.773                 & 6.146                 & 6.847                  &   4.420                     & \textbf{3.697}              \\
        Dinda's &            & MAE  & 0.071                     & \underline{0.041}       & 0.078                     & 0.053                   & 0.149                     & 0.057                     &  0.061                         & 0.046                 & 0.049                 & 0.059                  &   0.054                     & \textbf{0.029}              \\ \cline{2-15} 
        Dataset &            & MAPE & 0.079                     & \underline{0.030}       & 0.046                     & 0.040                   & 0.079                     & 0.032                     &  0.769                         & 0.035                 & 0.043                 & 0.058                  &   0.065                     & \textbf{0.018}              \\
                & Heavy      & MSE  & 17.925                    & \underline{6.837}       & 15.853                    & 9.202                   & 23.817                    & 8.729                     &  15.504                        & \textbf{5.635}        & 7.325                 & 13.274                 &   8.760                     & 6.124                   \\
                &            & MAE  & 0.104                     & \underline{0.040}       & 0.058                     & 0.051                   & 0.097                     & 0.041                     &  0.123                         & 0.046                 & 0.057                 & 0.076                  &   0.083                     & \textbf{0.025}              \\ \hline
                &            & MAPE & 0.238                     & \underline{0.237}       & 0.296                     & 0.278                   & 0.488                     & 0.290                     &  0.951                         & 0.207                 & 0.191                 & 0.161                  &   \textbf{0.076}            & 0.154            \\
                & Overall    & MSE  & 3.627                     & 3.282                   & 3.909                     & 4.950                   & 8.021                     & \underline{2.350}         &  2.012                         & 3.144                 & 2.974                 & 3.101                  &   \textbf{0.552}            & 1.757            \\
        SMD     &            & MAE  & 0.023                     & \underline{0.022}       & 0.025                     & 0.028                   & 0.046                     & 0.023                     &  0.040                         & 0.020                 & 0.019                 & 0.018                  &   0.015                     & \textbf{0.014}            \\ \cline{2-15} 
        Dataset &            & MAPE & 0.163                     & 0.166                   & \underline{0.154}         & 0.316                   & 0.367                     & 0.157                     &  0.689                         & 0.138                 & 0.129                 & 0.158                  &   0.204                     & \textbf{0.112}            \\
                & Heavy      & MSE  & 10.622                    & 11.024                  & 11.912                    & 17.587                  & 27.465                    & \underline{5.531}         &  3.652                         & 10.941                & 9.751                 & 10.399                 &   \textbf{4.138}            & 6.729                     \\
                &            & MAE  & 0.045                     & 0.041                   & 0.046                     & 0.068                   & 0.095                     & \underline{0.034}         &  0.056                         & 0.042                 & 0.038                 & 0.040                  &   0.055                     & \textbf{0.033}            \\ \hline
                &            & MAPE & 0.095                     & 0.058                   & 0.157                     & 0.086                   & 0.037                     & \underline{0.025}         &  0.135                         & 0.050                 & 0.027                 & 0.126                  &   0.017                     & \textbf{0.016}            \\
                & Overall    & MSE  & 0.489                     & \underline{0.408}       & 12.721                    & 0.880                   & 1.173                     & 0.510                     &  14.168                        & 0.359                 & 0.341                 & 0.975                  &   0.410                     & \textbf{0.323}            \\
        Alibaba &            & MAE  & 0.017                     & \underline{0.015}       & 0.058                     & 0.023                   & 0.026                     & 0.016                     &  0.056                         & 0.014                 & 0.013                 & 0.024                  &   0.014                     & \textbf{0.013}            \\ \cline{2-15} 
        Memory  &            & MAPE & 0.023                     & 0.027                   & 0.039                     & \underline{0.022}       & 0.032                     & 0.030                     &  0.069                         & 0.019                 & 0.019                 & 0.030                  &   0.019                     & \textbf{0.019}            \\
                & Heavy      & MSE  & \underline{0.673}         & 0.919                   & 2.420                     & 0.695                   & 1.418                     & 1.182                     &  8.624                         & 0.566                 & 0.542                 & 1.114                  &   0.598                     & \textbf{0.509}            \\
                &            & MAE  & 0.020                     & 0.024                   & 0.032                     & \underline{0.019}       & 0.029                     & 0.027                     &  0.065                         & 0.018                 & 0.018                 & 0.028                  &   0.018                     & \textbf{0.017}            \\ \hline
                &            & MAPE & 0.215                     & 0.165                   & 0.272                     & 0.190                   & 0.258                     & \underline{0.158}         &  0.266                         & \textbf{0.139}        & 0.158                 & 0.150                  &   0.153                     & 0.142                     \\
                & Overall    & MSE  & 4.996                     & \underline{4.809}       & 12.638                    & 6.097                   & 9.954                     & 6.204                     &  7.033                         & 4.786                 & 4.812                 & 4.784                  &   6.039                     & 4.758                     \\
        Alibaba &            & MAE  & 0.053                     & \underline{0.050}       & 0.090                     & 0.060                   & 0.073                     & 0.056                     &  0.060                         & 0.050                 & 0.051                 & 0.050                  &   0.054                     & \textbf{0.049}            \\ \cline{2-15} 
        CPU     &            & MAPE & 0.237                     & \underline{0.193}       & 0.343                     & 0.247                   & 0.381                     & 0.252                     &  0.342                         & 0.178                 & 0.177                 & 0.182                  &   0.176                     & \textbf{0.143}            \\
                & Heavy      & MSE  & \underline{17.548}        & 17.742                  & 29.493                    & 18.848                  & 32.271                    & 22.226                    &  38.502                        & 16.594                & 16.321                & 16.405                 &   14.090                    & \textbf{11.634}           \\
                &            & MAE  & \underline{0.100}         & \underline{0.100}       & 0.150                     & 0.105                   & 0.145                     & 0.115                     &  0.153                         & 0.097                 & 0.095                 & 0.095                  &   0.087                     & \textbf{0.082}            \\ \hline
    \end{tabular}%
}

\end{table*}

\subsection{Prediction Accuracy}
We summarize the performance of all the methods in Tab. \ref{Result}. We highlight the highest accuracy in boldface. Besides, if PePNet achieves the highest accuracy, we highlight the best one among its peers with underlines. 

As shown in Tab. \ref{Result}, PePNet achieves the best heavy-workload prediction accuracy as well as overall prediction accuracy on all three workload datasets, with few exceptions. Comparing the prediction accuracy on all three datasets, PePNet works the best on long-tail distributed Dinda's dataset, while less advantageous on the relatively even-distributed Alibaba2018-CPU dataset.

\subsection{Time Overhead}
We use an Intel(R) Xeon(R) CPU E5-2620 @ 2.10GHz CPU and a K80 GPU to record the time spent on training and inference for all methods. We show the time overhead of all the methods in Fig. \ref{fig:Time}. PePNet brings just an acceptable extra time overhead compared to the most efficient network (GRU) and greatly improves the overall and heavy-workload prediction accuracy, especially on Dinda's dataset (40.8\% for MAPE of heavy-workload prediction). 

\subsection{Hyperparameter Sensitivity}
\textbf{Impact of input length and prediction length. }
We use grid search~\cite{zhang2018online} to explore the impact of $\bar{X}$'s length and the impact of $y$'s length on PePNet's performance. We show the result in Fig. \ref{fig:cpuO}. We perform experiments with Cartesian combinations of input lengths from 30 to 80 and prediction lengths from 2 to 12. Generally, the error grows higher gently when predicting length increases. The best input length for different datasets is different. It is found the best input length for Alibaba2018-Memory and Alibaba2018-CPU is 50, while the best one for Dinda's dataset is 30.

\emph{Dinda's dataset.}
For overall workload prediction, MAE rises slowly as prediction length increases and MAE is stable for any input length. For heavy-workload workload prediction, MAE is more stable for prediction length when the input length is longer. As the length of prediction grows, MAE for overall accuracy increases no more than 0.051 at each input length, while MAE of heavy workload prediction increases no more than 0.062.

\emph{Alibaba2018-Memory.} The long-term dependency of Alibaba's memory usage is weak, thus when the input length is too big, increasing it increases the noise level, which could corrupt the performance of PePNet. Thus, when the input length is greater than 50, PePNet's MAE is unstable. But when the input length is less than 50, its performance is stable, and the prediction error only increases slowly with the prediction length. For overall memory usage prediction, MAE with a prediction length of 12 is only 0.006 larger than MAE with a prediction length of 2. For heavy-workload memory usage prediction, MAE with a prediction length of 12 is only 0.012 larger than that with a prediction length of 2. 

\emph{Alibaba2018-CPU.} Long-term dependency is more pronounced in CPU usage. Thus, when we increase the input length, the performance of PePNet is always stable. For overall CPU usage prediction, MAE with a prediction length of 12 is only 0.016 larger than that with a prediction length of 2. For heavy-workload CPU usage prediction, MAE with a prediction length of 12 is only 0.034 larger than that with a prediction length of 2.

\textbf{Impact of $\gamma$. }
We explore the impact of $\gamma$ on three datasets, as shown in Fig. \ref{fig:Gamma}. 
There are slight fluctuations of overall accuracy for different $\gamma$.
Besides, it is consistent with our prior theoretical analysis that when $\gamma$ is smaller the heavy-workload accuracy is higher. It is also intuitive that the heavy-workload MAE of Dinda's dataset dips most sharply, as Dinda's dataset is long-tail distributed. The extremely heavy workload in the long tail greatly reduces the prediction accuracy. But as $\gamma$ gets smaller, more attention is put on heavy workload, during the process of training. According to our experimental results, setting $\gamma$ as the value indefinitely approaching zero can lead to the best accuracy. As analyzed before, when $\gamma$ indefinitely approaches zero, AHLF only optimizes the part with the highest prediction error. In this situation, PePNet also achieves high prediction accuracy without overall accuracy degradation, which verifies that AHLF's use helps PePNet to avoid overly emphasizing sporadic patterns.

\subsection{Ablation Experiment}
We validate the effect of Periodicity-Perceived Mechanism and heavy-workload-focused loss function by comparing the performance of PePNet with those of PePNet$^-$ and PePNet$^\dagger$.
We show the performance of these models in Tab. \ref{Result}. Overall, Periodicity-Perceived Mechanism contributes more to the prediction accuracy than the AHLF. But the latter can be used to improve the accuracy of periodic and aperiodic time series, while Periodicity-Perceived Mechanism can only promote the accuracy of periodic and lax-periodic data. Furthermore, we test the prediction accuracy of PePNet on aperiodic data.

\textbf{Performance of Periodicity-Perceived Mechanism.} 
Dinda's dataset and SMD dataset exhibit stricter periodicity than Alibaba2018 does \cite{zhu2019novel}. Accordingly, the mechanism promotes the prediction accuracy most on Dinda's dataset and SMD dataset.
For datasets with lax periodicities, such as Alibaba2018-Memory, the mechanism also promotes overall and heavy-workload prediction accuracy.
As for Alibaba2018-CPU, which has no significant periodicity on light load but has more significant periodicity on heavy workload, the mechanism can promote the heavy-workload prediction accuracy, while the performance of PePNet is close to the one of PePNet$^-$ on light workload. This observation verifies that the mechanism has little negative impact on aperiodic data.

\textbf{Performance of AHLF.}
AHLF's use can improve both overall and heavy-workload prediction accuracies on all three datasets, compared with PePNet$^\dagger$. Except MAPE for heavy-workload prediction accuracy on memory usage of Alibaba2018, all of the evaluation metrics of PePNet are better than that of PePNet$^\dagger$. As AHLF apparently reduces the error for extremely heavy workload in the long tail, which is also seen in Fig. \ref{fig:Gamma}, its use improves the accuracy for Dinda's dataset and SMD dataset more that of Alibaba2018. 

\begin{table}[]
    \centering
    \renewcommand\arraystretch{1.2}
    \setlength\tabcolsep{0.5pt}
    \caption{\label{aperColl}Compare the performance of PePNet on aperiodic and periodic data. The MSE are multiplied by 1000. The MAE are multiplied by 100.}

    \begin{tabular}{c|c|ccc|ccc}
        \hline
        \multicolumn{1}{l|}{\multirow{2}{*}{}} & \multirow{2}{*}{Periodicity} & \multicolumn{3}{c|}{Overall Metric} & \multicolumn{3}{c}{Heavy-workload} \\ \cline{3-8} 
        \multicolumn{1}{l|}{}                  &                              & MAPE        & MSE       & MAE       & MAPE        & MSE        & MAE        \\ \hline
        \multirow{3}{*}{Ali18-Mem}        & Periodic                     & 0.024       & 0.08      & 0.47      & 0.018       & 0.17       & 0.95       \\
                                               & Aperiodic                    & 0.015       & 0.35      & 1.35      & 0.019       & 0.53       & 1.78       \\
                                               & Both                         & 0.016       & 0.32      & 1.27      & 0.019       & 0.51       & 1.74       \\ \hline
        \multirow{3}{*}{Ali18-CPU}           & Periodic                     & 0.133       & 2.81      & 3.79      & 0.117       & 5.44       & 5.16       \\
                                               & Aperiodic                    & 0.166       & 8.01      & 6.87      & 0.163       & 16.1       & 10.2       \\
                                               & Both                         & 0.142       & 4.76      & 4.94      & 0.143       & 12.4       & 8.21       \\ \hline
    \end{tabular}%

\end{table}

\textbf{Performance of PePNet on aperiodic data.}
There is a major concern about whether PePNet still works well on aperiodic data. To test the prediction accuracy on aperiodic data, we divide Alibaba2018-CPU and Alibaba2018-Memory by periodicity and collect the prediction accuracy for periodic data and aperiodic data, respectively.
In Tab. \ref{aperColl}, there is prediction accuracy for periodic data and aperiodic data. For convenience, the overall accuracy of both periodic data and aperiodic data is also listed at the bottom. On the whole, the accuracy for periodic data is slightly higher than the overall accuracy, while the accuracy for aperiodic data is slightly lower than the overall. However, there is a strange phenomenon MAPE of periodic data in Alibaba2018-Memory is bigger than the overall accuracy, while its MAE and MSE are much lower than the overall accuracy. That is because, in Alibaba2018-Memory dataset, the periodic workload is much smaller than the aperiodic one. Thus, even slight errors in periodic data prediction tend to become much larger after the division in the computation of MAPE. 

\section{Related Work}
Time series prediction aims to predict future time series based on the observed workload data. As an application branch of time series prediction, workload prediction aims to predict future workload based on observed ones. 

Existing studies have studied a wide range of workload prediction tasks, including GPU workloads \cite{hu2021characterization}, CPU utilization \cite{xue2022meta}, virtual desktop infrastructure pool workloads \cite{zhang2019cafe}, disk states \cite{de2020deep}, and database query arrival rates \cite{ma2018query}. These workload predictions are typically conducted to improve cloud system performance by optimizing the strategies of cloud resource allocation \cite{chen2021graph} and autoscaling \cite{xue2022meta}.

In addition, several studies have proposed integrated frameworks that combine prediction models with other functional modules. For instance, Uncertainty-Aware Heuristic Search \cite{luo2020intelligent} integrates workload prediction with Bayesian optimization to explicitly model uncertainty and guide virtual machine provisioning strategies. SimpleTS \cite{yao2023simplets} introduces an automated framework that selects appropriate prediction models for specific time series. However, these approaches primarily emphasize overall prediction accuracy, while largely overlooking accuracy under high-load conditions.

%\subsection{Periodicity Information Extraction}
%In the field of traffic forecasting, there is also significant periodicity and we summarize some recent popular methods below.
%There are several mechanisms to fuse periodic information. For example, Guo et al. propose a novel attention-based spatial-temporal graph convolution network \cite{guo2019attention}; Lv et al. use a fully connected layer and take advantage of both CNN and RNN to deal with periodic time series \cite{lv2018lc}; Chen et al. propose Hop Res-RGNN to deal with periodic patterns \cite{chen2019gated}.
%However, these methods do not consider the lax periodicity of time series.
%Yao et al. propose an attention mechanism to tackle periodic shift \cite{yao2019revisiting}. 
%These studies use the priori knowledge of the daily and weekly periodicity of traffic loads. 
%However, in the scenario of service load prediction, there is usually no priori knowledge about the periodicity. Besides, the periodicity of the time series in these studies is fixed, while the periodicities of workload for different machines are variable. 

\section{Conclusion}
In this paper, we study how to improve overall workload prediction accuracy as well as heavy-workload one and propose PePNet. PePNet makes use of short-term dependent information, long-term tendency information and periodic information.
Within PePNet, we propose two mechanisms for better prediction accuracy of workloads: 1) Periodicity-Perceived Mechanism to guide heavy-workload prediction, which can mine the periodic information and adaptively fuse periodic information for periodic, lax periodic and aperiodic time series; and 2) AHLF to offset the negative effect of data imbalance. We also provide theoretical support for the above design by:  1) providing a theoretically proven error bound of periodic information extracted by \emph{Periodicity-Perceived Mechanism}; and 2) providing an automatic hyperparameter determination method for periodicity threshold $\mathcal{T}$. Compared with existing methods, extensive experiments conducted on Alibaba2018, Dinda's dataset and SMD dataset demonstrate that PePNet improves MSE for overall workload prediction by 11.8\% on average.
Especially, PePNet improves MSE for heavy workload prediction by 21.0\% on average.

\appendices
\section{Proof of Theorem 1}
\label{app:theo1}
According to the proposed periodicity-mining algorithm, when \mn\ finds $k$, we have:
\begin{equation}
    \frac{E[a_{\tilde{t}}a_t]}{\sigma_{\tilde{t}}\sigma_{t}}>\mathcal{T}
    \label{cond1}
\end{equation}
\begin{equation}
    E[a_{\tilde{t}}a_t]>\mathcal{T}* \sigma_{\tilde{t}}\sigma_{t}
    \label{cond2}
\end{equation}
where $a_{i}=X_{i}-\mu_{i}$, $\tilde{t}=t-k$. $E[(X_{\tilde{t}}-X_t)^2]$ can be transformed to:
\begin{equation}
    E[(X_{\tilde{t}}-X_t)^2] = E[a_{\tilde{t}}^2+a_t^2-2a_{\tilde{t}}a_t+2\Delta\mu\Delta a+\Delta\mu^2]
    \label{cond3}
\end{equation}
Then, we substitute Eq.~(\ref{cond2}) into Eq.~(\ref{cond3}) to obtain:
\begin{equation}
    E[(X_{\tilde{t}}-X_t)^2] < E[a_{\tilde{t}}^2]+E[a_t^2]-2*\mathcal{T} \sqrt{E[a_{\tilde{t}}^2]E[a_t^2]}+\Delta\mu^2
    \label{cond4}
\end{equation}
Eq.~(\ref{cond4}) can be further reduced into: 
\begin{equation}
    E[(X_{\tilde{t}}-X_t)^2] < \Delta\sigma^2+\Delta\mu^2+2(1-\mathcal{T})\sigma_{\tilde{t}}\sigma_t
    \label{cond5}
\end{equation}
Hence, the first part of Theorem 1 is proven. 

When the time series is stationary, $\Delta\sigma^2$ and $\Delta\mu^2$ are zero. Then the upper bound can be reduced to $2(1-\mathcal{T})\sigma_{t-k}\sigma_{t}$.
% you can choose not to have a title for an appendix
% if you want by leaving the argument blank

\section{Proof of Theorem 2}
\label{app:theo2}
The derivation of $l(Y,y)$ is:
\begin{equation}
    \label{prove1}
    \frac{\partial l(Y,y)}{\partial \mathcal{P}}=\frac{\sum_t^T[exp(\frac{g(t,\mathcal{P})}{\gamma})\frac{\partial g(t,\mathcal{P})}{\partial \mathcal{P}}]}{\sum_t^T exp(\frac{g(t,\mathcal{P})}{\gamma})}
\end{equation}
Then, both numerator and dominator on the right-hand side are divided by $\exp(\frac{g(t^*,\mathcal{P})}{\gamma})$ simultaneously to obtain: 
\begin{equation}
    \label{prove2}
    \begin{aligned}
        &\mathop{\lim}_{\gamma\rightarrow 0}\frac{\partial l(Y,y)}{\partial \mathcal{P}} \\
        &=\mathop{\lim}_{\gamma\rightarrow 0}\frac{\frac{\partial g(t^*,\mathcal{P})}{\partial \mathcal{P}}+\sum_{t,t\neq t^*}^T [exp(\frac{g(t,\mathcal{P})-g(t^*,\mathcal{P})}{\gamma}) \frac{\partial g(t,\mathcal{P})}{\partial \mathcal{P}}]}
    {1+\sum_{t,t\neq t^*}^T exp(\frac{g(t,\mathcal{P})-g(t^*,\mathcal{P})}{\gamma})}
    \end{aligned}
\end{equation}
$g(t^*,\mathcal{P})$ is surely bigger than $g(t,\mathcal{P}), \forall t\neq t^*$, as the definition of $t^*$ suggests. Thus, $g(t,\mathcal{P})-g(t^*,\mathcal{P})<0, \forall t\neq t^*$. When $\gamma (\gamma>0)$ infinitely approaches zero, $exp(\frac{g(t,\mathcal{P})-g(t^*,\mathcal{P})}{\gamma})$ infinitely approaches zero as well. In this way, we have:
\begin{equation}
    \begin{aligned}
        &\mathop{\lim}_{\gamma\rightarrow 0}\frac{\frac{\partial g(t^*,\mathcal{P})}{\partial \mathcal{P}}+\sum_{t,t\neq t^*}^T [exp(\frac{g(t,\mathcal{P})-g(t^*,\mathcal{P})}{\gamma}) \frac{\partial g(t,\mathcal{P})}{\partial \mathcal{P}}]}
    {1+\sum_{t,t\neq t^*}^T exp(\frac{g(t,\mathcal{P})-g(t^*,\mathcal{P})}{\gamma})}\\
    &=\frac{\partial g(t^*,\mathcal{P})}{\partial \mathcal{P}} 
    \end{aligned}
\end{equation}

%{\appendices
%\section*{Proof of the First Zonklar Equation}
%Appendix one text goes here.
% You can choose not to have a title for an appendix if you want by leaving the argument blank
%\section*{Proof of the Second Zonklar Equation}
%Appendix two text goes here.}

\bibliographystyle{IEEEtran}
\bibliography{sample-base}

\end{document}